\documentclass[a4paper,12pt]{article}
\usepackage{amsmath,amsfonts,amssymb,cite}

\begin{document}

\begin{center}
{\Large\bf The effect of higher dimensional QCD operators on the spectroscopy of bottom-up holographic models}
\end{center}

\begin{center}
{\large S. S. Afonin%\footnote{E-mail: \texttt{s.afonin@spbu.ru}}
}
\end{center}

\begin{center}
{\it Saint Petersburg State University, 7/9 Universitetskaya nab.,
St.Petersburg, 199034, Russia}
%\\ E-mail: %\texttt{s.afonin@spbu.ru}
\end{center}

\begin{abstract}
Within the bottom-up holographic approach to QCD, the highly excited hadrons are identified
with the bulk normal modes in the fifth "holographic" dimension. We show
that additional states in the same mass range can appear also from taking into consideration
the 5D fields dual to higher dimensional QCD operators. The possible effects of
these operators were not taken into
account in almost all phenomenological applications. Using the scalar case as the simplest
example, we demonstrate that the additional higher dimensional operators lead to a large
degeneracy of highly excited states in the Soft Wall holographic model while in the
Hard Wall holographic model, they result in a proliferation of excited states. The considered model
can be viewed as the first analytical toy-model predicting a one-to-one mapping of the
excited meson states to definite QCD operators to which they prefer to couple.
\end{abstract}

\bigskip

The bottom-up holographic models for strong interactions enjoyed a surprising phenomenological
success in description of various experimental data. The most known models in this area
are the Hard Wall (HW)~\cite{son1,pom} and Soft Wall (SW)~\cite{son2}
holographic models. By construction, it turned out to be convenient to describe
the physics of chiral symmetry breaking and pion formfactors
within the the framework of HW model~\cite{pom} (see Ref.~\cite{br3} for a review) while the SW model
is well accommodated for description of linear Regge and radial trajectories
which were observed in light mesons~\cite{ani,bugg,klempt,afonin}.
Since the publication of those first papers introducing the bottom-up AdS/QCD approach,
a dramatic development has happened in the field that now includes several hundred papers.
To mention a few, numerous extensions of SW model were proposed aimed at improving various aspects
in phenomenological description of hadron spectroscopy and chiral symmetry
breaking~\cite{bottom-up2,bottom-up3,bottom-up4,forkel,bottom-up5,bottom-up7,bottom-up8,bottom-up9,bottom-up10,bottom-up11,zuo,bottom-up12,genSW,UV,UV3,nonlSW}.
The area of research included the studies of glueball sector~\cite{holSR,holSR2}, relations with light-front QCD (reviewed in~\cite{br3}) and with QCD sum rules~\cite{holSR3,holog2010},
important subjects of hadron form-factors (for a review see Refs.~\cite{br3,schmidt}) and gluon parton densities~\cite{Lyubovitskij}. The bottom-up
holographic study of QCD phase diagram was initiated in Ref.~\cite{Herzog} and gave rise to an interesting spin-off direction (for a review see Refs.~\cite{contr,AK}).

The observed hadrons are composite excited states in QCD, hence, as any composite excitations in a field theory,
theoretically they should be identified with the poles of some correlation functions.
The simplest and most important class of such functions is given by the two-point correlators $\langle J(x)J(y)\rangle$, where $J(x)$ represents
a local QCD operator that interpolates the hadron states under consideration. The mass spectrum of these states
appears from the poles of the corresponding two-point correlation function. It is generally believed that the meson masses are only slightly
(at the level of 10\%) changed when the large-$N_c$ limit of QCD is taken~\cite{hoof,wit} and the idea to calculate them in this limit
became extremely fruitful. The correlator $\langle J(x)J(y)\rangle$ in the large-$N_c$ limit is a meromorphic function and the higher $n$-point functions vanish~\cite{wit}.
This entails that in the limit $N_c\rightarrow\infty$,
$\langle J(x)J(y)\rangle$ has the structure of sum over infinite number
of pole terms corresponding to contributions of infinitely narrow mesons with quantum numbers determined by the operator $J$,
\begin{equation}
\label{sum}
\langle J(q)J(-q)\rangle\sim\sum_{n=0}^\infty\frac{Z_n}{q^2-m_n^2},
\end{equation}
in the momentum space, where contact terms needed for regularization are omitted.

The limit $N_c\rightarrow\infty$ is inherent in the very nature of holographic duality~\cite{witten,gub}.
The holographic method provides a definite prescription
how the correlation functions can be extracted from a dual theory~\cite{witten,gub}.
This prescription implies (due to the strong-weak duality) that the two-point correlation functions of 4D gauge theory follow from the quadratic
part of a dual theory in 5D Anti-de Sitter (AdS$_5$) space or asymptotically AdS$_5$ space, i.e. from the
free part of 5D action. The given property is exploited in constructing phenomenological bottom-up AdS/QCD
models --- in the first approximation, the meson spectrum should be given by the formally free quadratic part of a putative 5D theory.
The absence of interaction parts agrees with the requirement that the higher $n$-point functions vanish in the large-$N_c$ limit~\cite{wit}.

On the other hand, if one studies only the mass spectrum, it is not necessary to calculate the two-point correlators~\eqref{sum}.
Due to the known mathematical theorem on representation of the Green function via an infinite set of eigenfunctions, the mass spectrum (representing the spectrum
of corresponding eigenvalues) can be found from normalizable solutions of equation of motion of 5D holographic model~\cite{son1,pom}.
We will follow this way in our analysis.

The phenomenological holographic models typically contain a minimal set of fields.
This, via the holographic duality~\cite{witten,gub}, corresponds to restriction to
QCD operators of minimal dimension at fixed spin and other quantum numbers~\cite{son2}. If we consider
calculations of excited hadron spectrum from first principles in lattice QCD, the inclusion
of higher dimensional QCD operators turns out to be indispensable~\cite{Dudek}. The question
appears on the effect of these additional operators on the spectroscopy of holographic models.
The purpose of the present Letter is to answer this question.

We will be interested in what happens conceptually, the phenomenological fits will be left aside.
A consideration of scalar case will be enough for our purposes as generalizations to non-zero spins
are straightforward and do not change our general results.

We first recall briefly which gauge invariant scalar currents can be constructed in QCD.
The first set of operators is built from the quark fields and covariant derivatives,
\begin{equation}
\label{h2}
S^{(k)}=\bar{q}D^{2k}q,\qquad k=0,1,2,\dots,
\end{equation}
where $D^2=D^\mu D_\mu$. The isospin and $\gamma_5$ matrices can be also inserted in~\eqref{h2} but
this will be not essential for our further discussions. The canonical dimensions of currents~\eqref{h2} are
\begin{equation}
\label{3b}
\Delta=3+2k.
\end{equation}
The next set is given by various mixed operators with dimensions~\eqref{3b} but in which some of covariant derivatives
in~\eqref{h2} are replaced by the gluon field strength $G_{\mu\nu}$. For instance, apart from the operator $\bar{q}D^2 q$
with $\Delta=5$, we can construct the operator $\bar{q}G^{\mu\nu}\gamma_\mu\gamma_\nu q$ of the same dimension.
Also, for $\Delta=7$, along with $\bar{q}D^4 q$ the operators $\bar{q} D^2 G^{\mu\nu}\gamma_\mu\gamma_\nu q$
and $\bar{q}G^2 q$ are possible.
Another example of mixed operators is given by the scalar operators interpolating hybrid states of the
kind
\begin{equation}
\label{h3}
S^{(k)}_\text{h}=\bar{q}\gamma^{\mu}G_{\mu\nu}D^{\nu}\dots q,
\end{equation}
where dots denote insertions of $D^{2k}$ or $G^{2+k}$.
They will have an even canonical dimension
\begin{equation}
\label{3c}
\Delta=6+2k.
\end{equation}
Finally, the pure gluon scalar operators of the kind
\begin{equation}
\label{h4}
S^{(k)}_\text{gl}=G^{2+k},\qquad k=0,1,2,\dots,
\end{equation}
can be built. Their canonical dimensions are
\begin{equation}
\label{3d}
\Delta=4+2k.
\end{equation}

The currents~\eqref{h2},~\eqref{h3} and~\eqref{h4} transform differently under linear chiral transformations.
In the two-flavor case, the chiral group is $SU_\text{L}(2)\times SU_\text{R}(2)$. Since
the Dirac spinor can be decomposed as $q=q_\text{L}+q_\text{R}$, where $q_\text{L,R}=\frac{1\mp\gamma_5}{2}q$,
these currents transform as $(1/2,1/2)$, $(1,0)+(0,1)$ and $(0,0)$ representations of the chiral group correspondingly.

The relations~\eqref{3b},~\eqref{3c} and~\eqref{3d} can be written as one relation
\begin{equation}
\label{3e}
\Delta=r+2(k+1), \qquad k=0,1,2,\dots,
\end{equation}
where $r$ is equal to 1, 4 or 2 depending on aforementioned representations of the chiral group to which the
corresponding interpolating currents belong.

The mesonic currents including more than two quark fields are suppressed in the large-$N_c$ limit of QCD~\cite{hoof,wit},
for which the holographic description is hoped to be applicable. The baryonic currents in this limit interpolate
heavy solitonic objects~\cite{wit} and will not be considered.

The twist-3 scalar current $S^{(0)}$ in~\eqref{h2} has been traditionally used for interpolation
of scalar mesons in QCD sum rules, lattice QCD and low-energy effective
field theories. It is natural to expect that the radially excited mesons are stronger coupled
to operators of higher dimensions of the kind $S^{(k)}$ and their various variants corresponding to
inclusion of $G_{\mu\nu}$. Loosely speaking, a physical motivation for this belief is that in highly excited states,
the quark fields should undergo a faster change in the coordinate space and that these states should include
more gluons.

Let us now consider the simplest version of the SW model for scalars which is defined
by the 5D action
\begin{equation}
\label{1}
S=\frac12c^2\int d^4\!x\,dz\sqrt{g}\,e^{-az^2}\!\left(\partial^M\!\Phi\partial_M\!\Phi-m_5^2\Phi^2\right),
\end{equation}
where $g=|\text{det}g_{MN}|$, $M,N=0,1,2,3,4$, $c$ is a normalization constant for the
scalar field $\Phi$, and the parameter $a$ (that can be both positive and negative)
dictates the mass scale in the model. The metric of background AdS$_5$ space is usually
parametrized by the Poincar\'{e} patch with the line element
\begin{equation}
\label{2}
g_{MN}dx^Mdx^N=\frac{R^2}{z^2}(\eta_{\mu\nu}dx^{\mu}dx^{\nu}-dz^2),\qquad z>0.
\end{equation}
Here $\eta_{\mu\nu}=\text{diag}(1,-1,-1,-1)$, $R$ is the radius of AdS$_5$ space
and $z$ represents the holographic coordinate.
At each fixed $z$ one has the metric of flat 4D Minkowski space.
According to the standard prescriptions of AdS/CFT correspondence~\cite{witten,gub}
the 5D mass $m_5$ is determined by the relation,
\begin{equation}
\label{3}
m_5^2R^2=\Delta(\Delta-4),
\end{equation}
where $\Delta$ means the scaling dimension of 4D operator dual to the corresponding 5D field on the UV boundary.
The minimal value of dimension for scalar operator in QCD is $\Delta=3$ (the current~\eqref{h2}).
But, as discussed above, QCD interpolating operators can have higher canonical dimensions
and we will be interested in this general situation.

Generally speaking, we have of course many scalar fields with different 5D masses ---
one for each value of $\Delta$ in~\eqref{3}. We display in the action~\eqref{1} just
a common term for each such field $\Phi_\Delta$. In addition, each $\Phi_\Delta$ should be split
into several fields according to their different $r$-value in~\eqref{3e} reflecting different flavor and gluon content,
$\Delta\rightarrow\Delta_r$. All these technical points are implied and, inasmuch as we wish to demonstrate
the main idea in the most compact way, they will not be important in
our further discussion of emergent degeneracy which will occur for each $r$ separately.

The mass spectrum of physical 4D normal modes can be found, as usual, from the equation of
motion accepting the 4D plane-wave ansatz
\begin{equation}
\Phi(x_\mu,z)=e^{ipx}\phi(z),
\end{equation}
with the on-shell condition $p^2=m^2$. The equation of motion is
\begin{equation}
\label{12}
\partial_z\left(\frac{e^{-az^2}}{z^3}\partial_z \phi_n\right)=\left(\frac{m_5^2R^2}{z^5}-\frac{m_n^2}{z^3}\right)e^{-az^2}\phi_n,
\end{equation}
where $\phi_n$ correspond to normalizable discrete modes we are looking for.
After the substitution $\phi_n=z^{3/2}e^{az^2/2}\psi_n$, the Eq.~\eqref{12}
transforms into a one-dimensional Schr\"{o}dinger equation
\begin{equation}
\label{6}
-\partial_z^2\psi_n+V(z)\psi_n=m_n^2\psi_n,
\end{equation}
with the potential
\begin{equation}
\label{14}
V(z)=a^2z^2+\frac{4+m_5^2R^2-1/4}{z^2}+2a.
\end{equation}

The Schr\"{o}dinger equation of the kind
\begin{equation}
-\psi''+\left[x^2+\frac{m^2-1/4}{x^2}\right]\psi=E\psi,
\end{equation}
has eigenvalues
\begin{equation}
\label{15b}
E_n=4n+2m+2,\qquad n=0,1,2,\dots,
\end{equation}
and normalized eigenfunctions
\begin{equation}
\label{26c}
\psi_n(x)=\sqrt{\frac{2n!}{(m+n)!}}\,e^{-x^2/2}x^{m+1/2}L_n^m\left(x^2\right),
\end{equation}
where $L_n^m(x)$ are associated Laguerre polynomials. In Quantum Mechanics, this equation is known as
a radial equation for a two-dimensional harmonic oscillator with orbital momentum $m$ if $m$ is integer.

The eigenvalues~\eqref{15b} yield the mass spectrum in our case,
\begin{equation}
\label{15}
m_n^2=2|a|\left(2n+1+\frac{a}{|a|}+\sqrt{4+m_5^2R^2}\right),\qquad n=0,1,2,\dots.
\end{equation}
Making use of Eq.~\eqref{3} in~\eqref{15} we obtain
\begin{equation}
\label{16}
m_n^2=2|a|\left(2n+\frac{a}{|a|}+\Delta-1\right).
\end{equation}
Now we substitute the relation~\eqref{3e} and get finally
\begin{equation}
\label{17}
m_n^2=4|a|\left(n+k+\frac{r+1}{2}+\frac{a}{2|a|}\right),\qquad n,k=0,1,2,\dots.
\end{equation}
The mass spectrum~\eqref{17} reveals a surprising property: The numbers $n$ and $k$ can be interchanged.
Physically this means that the higher dimensional interpolating operators bring no states
with masses different from those predicted by the traditional SW model.

A difference appears, however, if we consider the normalized eigenfunctions corresponding to the discrete spectrum~\eqref{17},
\begin{equation}
\label{26b}
\phi_n=\sqrt{\frac{2n!}{(r+2k+n)!}}\,e^{(a-|a|)z^2/2}\left(|a|z^2\right)^{1+k+r/2}L_n^{r+2k}\left(|a|z^2\right),
\end{equation}
It is seen that the numbers $n$ and $k$ are not completely interchangeable
in the radial wave function: While the large $z$ asymptotics depends on the sum $n+k$ (because $L_n^\alpha(x)\sim x^n$ at large $x$),
the number of zeros is controlled by $n$ only (as the polynomial $L_n^\alpha(x)$ has $n$ zeros).

We arrive thus at the conclusion that the inclusion of higher dimensional operators into the SW model
leads to a degeneracy of high radial excitations --- each of them represents a mixture of several states
of equal energy for which the sum $n+k$ is fixed. This should have a dramatic effect on holographic calculation of
formfactors which are given by overlapping integrals in $z$ of eigenfunctions with electromagnetic external current~\cite{br3}.
The degeneracy would make such a calculation less predictive because one needs to assume a relative weight of each state
at fixed energy. On the other hand, one may expect that the contribution to the overlapping integral will be the larger
the less oscillating is the wave function in holographic coordinate. Since the number of zeros is controlled by $n$,
the dominance is expected for the state with $n=0$, i.e. for the state corresponding to maximal $k$ in the sum $n+k$.
It means that the state interpolated by QCD operator of maximal allowed dimension is expected to dominate. The given observation
could shed some light on connection between higher dimensional QCD operators and high radial excitations of hadrons ---
a vital problem for lattice calculations of excited hadron spectrum~\cite{Dudek}.

The inclusion of higher dimensional operators into the HW holographic model~\cite{son1,pom} is straightforward ---
we simply set $a=0$ in the equation of motion~\eqref{12} and impose the infrared
cutoff $z_0$. The normalizable solution for discrete modes is then given by
\begin{equation}
\label{27}
\phi\sim z^{2}J_{\Delta-2}(mz),
\end{equation}
where $J_\alpha(x)$ is a Bessel function of the first kind. In the original papers~\cite{son1,pom},
the discrete spectrum was dictated by the Dirichlet boundary condition,
\begin{equation}
\label{28}
\partial_z\phi(m_nz_0)=0.
\end{equation}
Using the known properties of Bessel functions, $2\partial_x J_\alpha=J_{\alpha-1}-J_{\alpha+1}$ and
$2\alpha J_\alpha/x=J_{\alpha-1}+J_{\alpha+1}$, the condition~\eqref{28} leads to the equation
\begin{equation}
\label{29}
\Delta J_{\Delta-3}(m_nz_0)=(\Delta-4)J_{\Delta-1}(m_nz_0).
\end{equation}
The relation~\eqref{3e} for $\Delta$ can be further substituted.

It is seen that the spectra of radial modes and modes given by higher dimensional
operators do not coincide. For instance, the lowest dimension of quark-antiquark
scalar operator is $\Delta=3$. Then the roots of Eq.~\eqref{29} are
$m_nz_0\approx\{2.7,5.7,\dots\}$. The next scalar operator with identical chiral
properties has dimension $\Delta=5$. The first root of Eq.~\eqref{29} is then
$m_nz_0\approx4.9$ which is not equal to the second root above. The same
goes on for other dimensions $\Delta$ and radial modes.
We obtain thus a proliferation of highly excited states instead of degeneracy.

It is not difficult to generalize our discussion to the case of non-zero spin.
One can show that all general conclusions will remain the same.
The corresponding details will be presented elsewhere.


\begin{thebibliography}{99}


\bibitem{son1} Erlich, J.; Katz, E.; Son, D.T.; Stephanov, M.A.
``QCD and a holographic model of hadrons,''
\emph{Phys. Rev. Lett.} \textbf{2005}, \emph{95}, 261602.

\bibitem{pom} Rold, L.D.; Pomarol, A.
``Chiral symmetry breaking from five dimensional spaces,''
\emph{Nucl. Phys. B} \textbf{2005}, \emph{721}, 79.

\bibitem{son2} Karch, A.; Katz, E.; Son, D.T.; Stephanov, M.A.
``Linear confinement and AdS/QCD,''
\emph{Phys. Rev. D} \textbf{2006}, \emph{74}, 015005.

\bibitem{br3}
Brodsky, S.J.; de Teramond, G.F.; Dosch, H.G.; Erlich, J.
``Light-Front Holographic QCD and Emerging Confinement,''
\emph{Phys. Rep.} \textbf{2015}, \emph{584}, 1.

\bibitem{ani} Anisovich, A.V.; Anisovich, V.V.; Sarantsev, A.V.
``Systematics of q anti-q states in the (n, M**2) and (J, M**2) planes,''
\emph{Phys. Rev. D} {\bf 2000}, \emph{62}, 051502(R).

\bibitem{bugg} Bugg, D.V.
``Four sorts of meson,''
\emph{Phys. Rep.} \textbf{2004}, \emph{397}, 257.

\bibitem{klempt} Klempt, E.; Zaitsev, A.
``Glueballs, Hybrids, Multiquarks. Experimental facts versus QCD inspired concepts,''
\emph{Phys. Rep.} \textbf{2007}, \emph{454}, 1.

\bibitem{afonin} Afonin, S.S.
``Implications of the Crystal Barrel data for meson-baryon symmetries,''
\emph{Mod. Phys. Lett. A} \textbf{2008}, \emph{23}, 3159.

\bibitem{bottom-up2}
Boschi-Filho, H.; Braga, N.R.F.; Carrion, H.L. ``Glueball Regge trajectories from gauge/string duality and the Pomeron,''
\emph{Phys. Rev. D} \textbf{2006}, \emph{73}, 047901.

\bibitem{bottom-up3}
Hirn, J.; Rius, N.; Sanz, V. ``Geometric approach to condensates in holographic QCD,''
\emph{Phys. Rev. D} \textbf{2006}, \emph{73}, 085005;
Ghoroku, K.; Maru, N.; Tachibana, M.; Yahiro, M. ``Holographic model for hadrons in deformed AdS(5) background,''
\emph{Phys. Lett. B} \textbf{2006}, \mbox{\emph{633}, 602.}

\bibitem{bottom-up4}
Cs\'{a}ki, C.; Reece, M. ``Toward a systematic holographic QCD: A Braneless approach,''
\emph{J. High Energy Phys. (JHEP)} \textbf{2007}, \emph{0705}, 062;
Shock, J.P.; Wu, F.; Wu, Y.-L.; Xie, Z.-F. ``AdS/QCD Phenomenological Models from a Back-Reacted Geometry,''
\emph{J. High Energy Phys. (JHEP)} \textbf{2007}, \emph{0703}, 064.

\bibitem{forkel}  Forkel, H; Beyer, M.; Frederico, T.
``Linear square-mass trajectories of radially and orbitally excited hadrons in holographic QCD,''
\emph{J. High Energy Phys. (JHEP)} \textbf{2007}, \emph{0707}, 077.

\bibitem{bottom-up5}
Batell, B.; Gherghetta, T. ``Dynamical Soft-Wall AdS/QCD,''
\emph{Phys. Rev. D} \textbf{2008}, \emph{78}, 026002;
Gherghetta, T.; Kapusta, J.I.; Kelley, T.M. ``Chiral symmetry breaking in the soft-wall AdS/QCD model,''
\emph{Phys. Rev. D} \textbf{2009}, \emph{79}, 076003.

\bibitem{bottom-up7}
de Paula, W.; Frederico, T.; Forkel, H.; Beyer, M.
``Dynamical AdS/QCD with area-law confinement and linear Regge trajectories,''
\emph{Phys. Rev. D} \textbf{2009}, \emph{79}, 075019.

\bibitem{bottom-up8}
Vega, A.; Schmidt, I. ``Hadrons in AdS/QCD correspondence,''
\emph{Phys. Rev. D} \textbf{2009}, \emph{79}, 055003.

\bibitem{bottom-up9}
Afonin, S.S.
``AdS/QCD models describing a finite number of excited mesons with Regge spectrum,''
\emph{Phys. Lett. B} \textbf{2009}, \emph{675}, 54;
``Regge spectrum from holographic models inspired by OPE,''
\emph{Phys. Lett. B} \textbf{2009}, \emph{678}, 477.

\bibitem{bottom-up10}
de Teramond, G.F.; Brodsky, S.J. ``Light-Front Holography: A First Approximation to QCD,''
\emph{Phys. Rev. Lett.} \textbf{2009}, \emph{102}, 081601;
Cherman, A.; Cohen, T.D.; Werbos, E.S. ``The Chiral condensate in holographic models of QCD,''
\emph{Phys. Rev. C} \textbf{2009}, \emph{79}, 045203;
Becciolini, D.; Redi, M.; Wulzer, A. ``AdS/QCD: The Relevance of the Geometry,''
\emph{J. High Energy Phys. (JHEP)} {\bf 2010}, \emph{1001}, 074.

\bibitem{bottom-up11}
Afonin, S.S; Pusenkov, I.V. ``The quark masses and meson spectrum: A holographic approach,''
\emph{Phys. Lett. B} \textbf{2013}, \emph{726}, 283.

\bibitem{zuo} Zuo, F. ``Improved Soft-Wall model with a negative dilaton,''
\emph{Phys. Rev. D} \textbf{2010}, \emph{82}, 086011;
Gutsche, T.; Lyubovitskij, V.E.; Schmidt, I.; Vega, A. ``Dilaton in a soft-wall holographic approach to mesons and baryons,''
\emph{Phys. Rev. D} \textbf{2012}, \emph{85}, 076003.

\bibitem{bottom-up12}
Gutsche, T.; Lyubovitskij, V.E.; Schmidt, I.; Vega, A.
``Chiral Symmetry Breaking and Meson Wave Functions in Soft-Wall AdS/QCD,''
\emph{Phys. Rev. D} \textbf{2013}, \emph{87}, 056001.

\bibitem{genSW} Afonin, S.S. ``Generalized Soft Wall Model,''
\emph{Phys. Lett. B} \textbf{2013}, \emph{719}, 399.

\bibitem{UV} Evans, N.; Tedder, A.
``Perfecting the Ultra-violet of Holographic Descriptions of QCD,''
\emph{Phys. Lett. B} \textbf{2006}, \emph{642}, 546;
Afonin, S.S. ``Low-energy holographic models for QCD,''
\emph{Phys. Rev. C} \textbf{2011}, \emph{83}, 048202.

\bibitem{UV3} Braga, N.R.F.; Contreras, M.A.M.; Diles, S.
``Decay constants in soft wall AdS/QCD revisited,''
\emph{Phys. Lett. B} \textbf{2016}, \emph{763}, 203;
``Holographic model for heavy-vector-meson masses,''
EPL \textbf{2016}, \emph{115}, 31002.

\bibitem{nonlSW}
Contreras, M.A.M.; Vega, A.
``Nonlinear Regge trajectories with AdS/QCD,''
\emph{Phys. Rev. D} \textbf{2020}, \emph{102}, 046007.

\bibitem{holSR} Forkel, H.
``Holographic glueball structure,''
\emph{Phys. Rev. D} \textbf{2008}, \emph{78}, 025001.

\bibitem{holSR2}
Colangelo, P.; Fazio, F.D.; Jugeau, F.; Nicotri, S.
``Investigating AdS/QCD duality through scalar glueball correlators,''
\emph{Int. J. Mod. Phys. A} \textbf{2009}, \emph{24}, 4177.
%[arXiv:0711.4747 [hep-ph]].
Colangelo, P.; Fazio, F.D.; Giannuzzi, F.; Jugeau, F.; Nicotri, S.
``Light scalar mesons in the soft-wall model of AdS/QCD,''
\emph{Phys. Rev. D} \textbf{2008}, \emph{78}, 055009.
%[arXiv:0807.1054 [hep-ph]].

\bibitem{holSR3} Jugeau, F.; Narison, S.; Ratsimbarison, H.
``SVZ 1/q$^2$-expansion versus some QCD holographic models,''
\emph{Phys. Lett. B} \textbf{2013}, \emph{722}, 111.
%[arXiv:1302.6909 [hep-ph]].

\bibitem{holog2010}
Afonin, S.S.
``Holographic like models as a five-dimensional rewriting of large-Nc QCD,''
\emph{Int. J. Mod. Phys. A} \textbf{2010}, \emph{25}, 5683;
``Note on Relation between Bottom-Up Holographic Models and Large- $N_c$ QCD,''
\emph{Adv. High Energy Phys.} \textbf{2017}, \emph{2017}, 8358473.

\bibitem{schmidt}
Gutsche, T.; Lyubovitskij, V.E.; Schmidt, I.
``Electromagnetic properties of the nucleon and the Roper resonance in soft-wall AdS/QCD at finite temperature,''
\emph{Nucl. Phys. B} \textbf{2020}, \emph{952}, 114934.

\bibitem{Lyubovitskij} Lyubovitskij, V.E.; Schmidt, I.
``Gluon parton densities in soft-wall AdS/QCD,''
\emph{arXiv} \textbf{2012}, arXiv:2012.01334.

\bibitem{Herzog}
Herzog, C.P.
``A Holographic Prediction of the Deconfinement Temperature,''
\emph{Phys. Rev. Lett.} \textbf{2007}, \emph{98}, 091601.
% [hep-th/0608151].

\bibitem{contr} Braga, N.R.F.; Contreras, M.A.M.; Diles, S.
``Holographic Picture of Heavy Vector Meson Melting,''
\emph{Eur. Phys. J. C} \textbf{2016}, \emph{76}, 598.

\bibitem{AK}Afonin, S.; Katanaeva, A. ``On holographic relation between radial meson trajectories and deconfinement temperature,''
\emph{Phys. Rev. D} \textbf{2018}, \emph{98}, 114027.

\bibitem{hoof}'t Hooft, G. ``A Planar Diagram Theory for Strong Interactions,''
\emph{Nucl. Phys. B}~\textbf{1974}, \emph{72}, 461.

\bibitem{wit} Witten, E. ``Baryons in the 1/n Expansion,''
\emph{Nucl. Phys. B}~\textbf{1979}, \emph{160}, 57.

\bibitem{witten} Witten, E. ``Anti-de Sitter space and holography,''
\emph{Adv. Theor. Math. Phys.} \textbf{1998}, \emph{2}, 253.

\bibitem{gub} Gubser, S.S.; Klebanov, I.R.; Polyakov, A.M. ``Gauge theory correlators from noncritical string theory,''
\emph{Phys. Lett. B} \textbf{1998}, \emph{428}, 105.

\bibitem{Dudek} Dudek, J.J.; Edwards, R.G.; Peardon, M.J.; Richards, D.G.; Thomas, C.E.  Hadron Spectrum Collaboration.
``Toward the excited meson spectrum of dynamical QCD,''
\emph{Phys. Rev. D} \textbf{2010}, \mbox{\emph{82}, 034508.}



\end{thebibliography}
\end{document}